\documentclass[12pt]{article}
\usepackage{axodraw}

\date{December 1995\\Revised April 1996\\hep-ph/9512428}
\author{
Juan Carlos D'Olivo \\
Instituto de Ciencias Nucleares\\
Universidad Nacional Aut\'{o}noma de M\'{e}xico\\
Apartado Postal 70-543, 04510 M\'{e}xico, D.F., M\'{e}xico\\
\and
Jos\'e F. Nieves\\
Laboratory of Theoretical Physics\\
Department of Physics, P. O. Box 23343\\
University of Puerto Rico\\
R\'{\i}o Piedras, Puerto Rico 00931-3343
}
\title{Chirality-preserving neutrino oscillations
in an external magnetic field}

\begin{document}

\maketitle

\begin{abstract}

Neutrinos propagating in  matter acquire
an effective electromagnetic vertex induced by their
weak interactions with the charged particles
in the background.  In the presence of an external
magnetic field the induced vertex affects the
flavor transformations of mixed neutrinos in a way that,
in contrast to the oscillations driven by
an intrinsic magnetic moment interaction,
preserve chirality.  We derive the evolution equation
for this case and discuss some of the physical
consequences in environments such as a
supernova. For small values of the square mass
difference the resonance  for neutrinos and
antineutrinos occur within regions which are close.  In that case,
the resonance condition
becomes independent
of the vacuum parameters and is approximately the same
for both.

\end{abstract}

As is now well known, under certain favorable conditions,
large transformations from one neutrino flavor into another
may take place in a medium, even for small neutrino mixing in
vacuum\cite{wolfenstein}. The matter effects
on the $\nu$ oscillations are taken into account
by means of a potential energy or an index of refraction
for each neutrino flavor, which can be calculated from
the background contributions to the neutrino
self-energy\cite{Notzold,PalPham,Nieves:neutrinos}.
In addition to the energy-momentum relation of the neutrinos,
also their electromagnetic
properties can be affected in an important
manner\cite{NievesPal1,DNP1,DNP2,semikoz,DNPnucerenk}.
In Ref.~\cite{DNP1} the general expressions for the
electromagnetic form factors of a neutrino in an electron gas
were calculated and, as an specific application,
the correction to the index of
refraction of a single neutrino in the presence
of an external magnetic field was determined.
In this paper we extend the study of Ref.~\cite{DNP1}
by considering the combined effects of neutrino mixing and
the external magnetic field. We derive the equation governing
the flavor evolution under such conditions and examine
the possible effects of strong magnetic fields on the neutrino
oscillations induced by their effective electromagnetic interaction.

It is worth stressing the following.
The possibility that neutrinos may have electromagnetic dipole moment
interactions
which change left-handed neutrinos into right-handed ones
can have important consequences in the context
of the solar-neutrino puzzle\cite{cisneros}, and the combined effect
of matter and magnetic fields on neutrino flavor
oscillations and spin precession has been studied\cite{okun}.
However, it is well known that the values of the neutrino magnetic moments
that are estimated in the Standard Model fall well
below the values for which their effect can be appreciable.
While there are schemes in which the values of the neutrino
magnetic moments lie in the relevant range, they require
ingredients that are not contained
in the standard Electroweak Theory.
On the contrary, the effect we are considering is quite
different.  The neutrino electromagnetic
form factors calculated in Ref.~\cite{DNP1} preserve
chirality and are present even in the Standard Model
with massless left-handed neutrinos.  Those induced
terms arise because of the interactions of neutrinos
with the particles in the background.  In the presence
of an external magnetic field the induced form factors,
instead of producing spin flip transitions,
contribute to the index of refraction and modify
the resonance condition for neutrino oscillations
in matter.

For simplicity we restrict ourselves to the case
of mixing between only two generations,
but the same approach can be
applied to more general cases.
Our starting point is  the self-energy
of the neutrinos, with momentum $k^\mu$, in the presence of a uniform
magnetic field. As shown in Ref.~\cite{DNP1}, it
can be written as
\begin{equation}
\label{sigmaeff}
\Sigma_{eff} = (a\rlap / k + b\rlap / u + c\rlap /\! B_{ext})L\,,
\end{equation}
where $u^\mu$ is the velocity four-vector of the medium, 
$L = (1 - \gamma_5)/2$, and
$B_{ext}^\mu = 1/2 \epsilon^{\mu\nu\alpha\beta}u_\nu F_{\alpha\beta}$,
where $F$ is the electromagnetic field tensor.
In what follows we work in the rest frame of
the background, where $u^\mu = (1,\vec 0)$
$k^\mu = (\omega,\vec\kappa)$, and $B_{ext}^\mu = (0,\vec B)$
with $\vec B$ being the external magnetic field.

The coefficients $a$, $b$, and $c$ are matrices in the neutrino
internal space and,
in a normal matter gas, composed of electrons, nucleons and their
antiparticles, they
are given by
\begin{eqnarray}\label{coefficients}
a & = &   0\,,\nonumber\\
b & = & \sqrt{2}G_FQ_Z  + \left(\begin{array}{cc}
b_e & 0\\
0 & 0
\end{array}\right)\nonumber\\
c & = & \sqrt{2}G_FQ_Z^\prime  + \left(\begin{array}{cc}
c_e & 0\\
0 & 0
\end{array}\right)
\end{eqnarray}
where, to order $G_F$,
\begin{eqnarray}\label{bece}
b_e & = & \sqrt{2}G_F(n_e - n_{{\overline e}})\nonumber\\
c_e & = & 2\sqrt{2}eG_F\int\frac{d^3p}{(2\pi)^3 2E}
\frac{d}{dE}(f_{-} - f_{+})\,,
\end{eqnarray}
with $e$ being the electron charge ($e < 0$).\footnote{We take
the opportunity to stress that in Ref.~\cite{DNP1}
the symbol $e$ stands everywhere for the electron charge
and not for its magnitude.}
We have introduced the
electron and positron distributions
\begin{equation}\label{fdistributions}
f_{\mp}(p) = \frac{1}{e^{\beta(E(p) \mp \mu)} + 1}\,,
\end{equation}
and the corresponding number densities
\begin{equation}\label{numberdensities}
n_{e,{\overline e}} = 2\int\frac{d^3p}{(2\pi)^3}f_{\mp}(p)\,.
\end{equation}

In Eq.~(\ref{coefficients}) we have denoted by
$Q_Z$ and $Q_Z^\prime$ the
contributions arising from the $Z$-diagram,
which are the same for all flavors and, in a
normal matter background, are irrelevant
for oscillations.
However, in environments like the early universe or the core of a
supernova, where the neutrinos represent an appreciable fraction
of the total density,
the neutral-current contributions to the potential energy
arising from the $\nu$-$\nu$ scattering are not in general
proportional to the unit matrix
and should be included in the analysis of the
resonant flavor transformations\cite{pantaleone}.
Several approaches exist to
describe the neutrino oscillations under such conditions,
including
a Boltzmann-type kinetic approach\cite{densitymatrix2}
based on a density matrix
formalism\cite{densitymatrix},
and a treatment based on the FTFT methods\cite{DNwolfeq}.
While we are aware that the extra contributions
from the $\nu$-$\nu$ background interactions
must in general be taken into account in a careful numerical study,
in what follows
we will not consider them further since they can be
added at any stage.  Of course, they will be important
in the particular context of the
supernova, for example\cite{fuller}.

\begin{figure}
\begin{center}
\begin{picture}(100,115)(0,0)
\ArrowLine(100,80)(0,80) \Text(50,86)[]{$e$}
\PhotonArc(50,80)(30,0,180){4}{8.5} 
\Text(50,120)[]{$W$}
\Photon(33,80)(33,40){3}{8.5}
\Photon(66,80)(66,40){3}{8.5}
\Text(48,10)[]{(a)}
\end{picture}
\hspace{1in}
\begin{picture}(100,135)(0,0)
\Oval(50,92)(30,20)(0) \Text(77,92)[]{$e$}
\ArrowLine(100,132)(50,122)
\ArrowLine(50,122)(0,132)
\Photon(62,22)(57,65){3}{8.5}
\Photon(38,22)(43,65){3}{8.5}
\Text(50,10)[]{(b)}
\end{picture}
\end{center}
\caption{\label{fig1}  Diagram (a) gives the dominant
contribution (of order $1/m_W^2$)
to the neutrino self-energy term of order $B^2$, 
arising from the $W$ exchange interaction.  In the local
limit of the $W$ propagator, this diagram is equivalent
to Diagram (b).}
\end{figure}
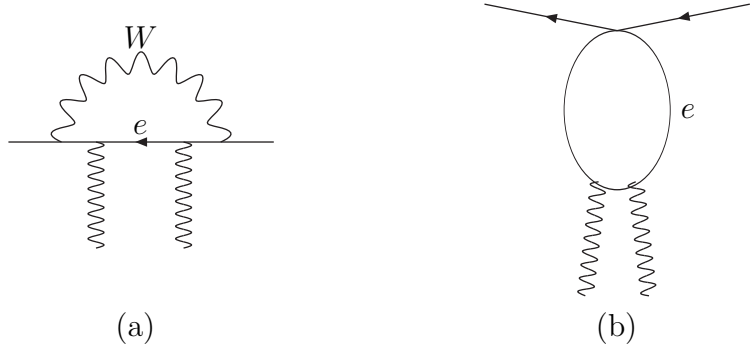

In general, the coefficients $a, b$ and $c$ are
functions of $B$. The calculation of Ref.\ \cite{DNP1}
corresponds to retaining
only the contribution to the coefficients that is
independent of $B$.  Since
we are envisaging a situation where $cB$ could be comparable
to $b$, it is pertinent to ask whether the second and higher
order terms in $B$ are important or not. 
In order to answer that, consider the contribution
to the neutrino self-energy arising from the diagram in which
two external $B$ lines are attached to the electron line
of the $W$-exchange loop diagram, as shown in
Fig.\ \ref{fig1}.

That gives is an additional contribution
to $b_e$ such that,
instead of the formula for $b_e$ given in Eq.\ (\ref{bece}),
the result is 
\begin{equation}\label{be2ndorder}
b_e = \sqrt{2}G_F\left(2\int\frac{d^3p}{(2\pi)^3}(f_{-} - f_{+})
+ b_2 B^2\right)\,,
\end{equation}
where $b_2$ is a coefficient independent of $B$.   The crucial
point now is to note that the term involving
the distribution functions $f_{\mp}$ cannot be identified
with the total densities as in Eq.\ (\ref{numberdensities}),
because the latter quantities must now be determined
to order $B^2$ also.  The diagram for the $B^2$
contribution
to the electron current density $\langle \overline e\gamma^\mu e\rangle$,
which can also be written as the trace of $S_F\gamma^\mu$ where
$S_F$ is the electron propagator,
is identical to Diagram (b) but with the external
neutrino lines removed.  Thus, by simple inspection
of the two diagrams
it is easy to recognize that the net number density
is given by precisely the same factor that
appears in Eq.\ (\ref{be2ndorder}),
\begin{equation}\label{netdensity2ndorder}
n_{-} - n_{+} = 2\int\frac{d^3p}{(2\pi)^3}(f_{-} - f_{+})
+ b_2 B^2 \,.
\end{equation}
This is now the relation that determines the
chemical potential in terms of the net number density,
which is the conserved quantity.  If we proceed to
eliminate the chemical potential in favor of
$n_{-} - n_{+}$, the result is that the formula
for $b_e$ reduces again to that in Eq.\ (\ref{bece}),
while $c_e$ acquires a dependence on $B$ through
the implicit dependence of the $f_{\mp}$ on
$B$, as a consequence of solving for the chemical
potential in Eq.\ (\ref{netdensity2ndorder}).
The same argument applies to higher order
terms in $B$.  This can be seen easily
by considering, for example, diagrams
with more $B$ lines attached to the electron,
and noting that they are equivalent
to the electron loop of Diagram (a) but
with the additional external $B$ lines.
All this amounts to the statement that, to
this order in $G_F$, the neutrino self-energy
is given by just
$\langle \overline e\gamma^\mu e\rangle\gamma_\mu L$.  Then,
according to Eq.\ (\ref{sigmaeff}), $b_e$
is determined by $\langle e^\dagger e\rangle$
which is just the net number density.
Thus, to summarize, the formula for $b_e$ in Eq.\ (\ref{bece})
is, to order $G_F$, independent of $B$.  On the other hand,
by adopting Eq.\ (\ref{numberdensities}) as the
relation between the chemical potential
and the net number density, we are neglecting
the dependence of $c_e$ on the magnetic field\footnote{For
the purpose of studying
this dependence, a better procedure is probably
to start from the expression for the neutrino
self-energy in terms of the electron propagator
in an external magnetic field, as recently carried
out by Elmfors, Grasso and Raffelt\cite{elmfors}}.

The coefficients $b, c$ in Eq.~(\ref{coefficients}) are written
in the flavor basis.  In this basis,
the equation that determines
the dispersion relation and wave functions of the
propagating modes is
\begin{equation}\label{fieldeq}
(\rlap / k - m - \Sigma_{eff})\psi = 0\,,
\end{equation}
where,
\begin{equation}\label{massmatrix}
m = U\left(\begin{array}{cc}
m_1 & 0\\
0 & m_2
\end{array}\right)U^\dagger\,,
\end{equation}
with $U$ being the matrix that relates the
fields of definite mass $\nu_{Li}$ to the
flavor fields $\nu_{L\alpha} = \sum_i U_{\alpha i}\nu_{Li}
\quad (\alpha = e,\mu)$.
In the Weyl representation
\begin{equation}\label{weylrep}
\psi_R = \left( \begin{array}{c}\xi\\0\end{array}\right)\,,
\ \ \ \ \
\psi_L = \left( \begin{array}{c}0\\\eta\end{array}\right)\,,
\end{equation}
and Eq.~(\ref{fieldeq}) becomes the set of coupled equations
\begin{eqnarray}\label{fieldeqweyl}
(\omega - b + \vec\sigma\cdot\vec\kappa - c\vec\sigma\cdot B)\eta -
m\xi & = & 0\,,\nonumber\\
(\omega - \vec\sigma\cdot\vec\kappa)\xi - m\eta & = & 0\,.
\end{eqnarray}
Using the second of these
to eliminate $\xi$ from the first one, yields
the following equation for $\eta$
\begin{equation}\label{etaeq}
\left[(\omega - b + \vec\sigma\cdot\vec\kappa
- c\vec\sigma\cdot\vec B) - (\omega + \vec\sigma\cdot\vec\kappa)
\frac{m^2}{\omega^2 - \kappa^2}\right]\eta = 0\,,
\end{equation}
while $\xi$ is then determined as
\begin{equation}\label{xi}
\xi = (\omega + \vec\sigma\cdot\vec\kappa)\frac{m}{\omega^2 - \kappa^2}\eta\,.
\end{equation}

We are interested in the solutions of Eq.~(\ref{etaeq}) with
positive energy, corresponding to the (neutrinos) particle
solutions.  In the absence of the magnetic field
they correspond to negative helicity spinors of the form
\begin{equation}
\eta = e_{1,2}\phi_{-}\,,
\end{equation}
where $\phi_\lambda$ is the Pauli spinor with definite
helicity, that satisfies
\begin{equation}
\left(\vec\sigma\cdot\hat\kappa\right)\phi_\lambda =
\lambda\phi_\lambda\,\quad (\lambda = \pm)\,.
\end{equation}
The $e_{1,2}$ are vectors in flavor space,
determined by solving
the eigenvalue problem
\begin{equation}
\label{eigenvalueeq}
He_i = \omega_i e_i\,,
\end{equation}
where
\begin{equation}
\label{ham}
H = \kappa + \frac{m^2}{2\kappa} + b\,.
\end{equation}
The expressions for $\omega_{1,2}$ and $e_{1,2}$
are given explicitly in
Eqs.~(A.10) and (A.14) of Ref.~\cite{DNwolfeq}.
To arrive at Eq.~(\ref{ham}), the substitution
\begin{equation}\label{approx}
(\omega + \vec\sigma\cdot\vec\kappa)
\frac{m^2}{\omega^2 - \kappa^2}\eta =
\frac{m^2}{\omega + \kappa}\eta\simeq
\frac{m^2}{2\kappa}\eta
\end{equation}
has been made.

In the presence of the magnetic field, the solutions
to Eq.~(\ref{etaeq})
do not correspond to purely negative helicity
spinors any longer because the matrices $\vec\sigma\cdot\vec B$
and $\vec\sigma\cdot\vec\kappa$ generally do not
commute.  Therefore, the solution must be sought
in the form
\begin{equation}\label{genansatz}
\eta = x\phi_{-} + x^\prime\phi_{+}\,,
\end{equation}
where $x, x^\prime$ are two-component vectors
in flavor space.
Substituting Eq.~(\ref{genansatz}) into Eq.~(\ref{etaeq})
we obtain two coupled equations for $x$ and $x^\prime$,
and from them is easy to verify
that $x^\prime\sim (c\vec B\cdot\hat\kappa/\kappa)x$.
Then, retaining terms that are at most linear in the small
quantities $b, c, m^2/2\kappa$ in the equation for $x$,
the dispersion relations
$\omega_{1,2}$
and the corresponding vectors $x = e_{1,2}$
are obtained by solving Eq.~(\ref{eigenvalueeq}), but with
the Hamiltonian
\begin{equation}\label{hamB}
H_B = \kappa + b - c\vec B\cdot\hat\kappa + \frac{m^2}{2\kappa}
\end{equation}
in place of $H$.

Imitating the arguments given in
Ref.~\cite{DNwolfeq}, we then arrive at the
following picture:  The Dirac wave function
for a relativistic left-handed neutrino
with momentum $\vec\kappa$ propagating in matter
in the presence of an external magnetic field
is, in the Weyl representation,
\begin{equation}\label{Diracwf}
\psi_L = e^{i\vec\kappa\cdot \vec x}
\left( \begin{array}{c}0\\\phi_{-}\end{array}\right)\chi(t)\,,
\end{equation}
where, in a homogenous medium,
$
\chi(t) = \sum_{i = 1,2} (e_i^\dagger\chi(0))e_i e^{-i\omega_i t}\,.
$
For an inhomogenous medium,
the flavor-space spinor $\chi(t)$
is the solution of
\begin{equation}
\label{wolfeq}
i\frac{d\chi}{dt} = H_B\chi\,,
\end{equation}
which is the extension of the MSW equation
to the situation we are considering.
The components $\chi_{e,\mu}$ of $\chi(t)$
give the amplitude to find the neutrino in the
corresponding state of definite flavor, at a distance
$r \simeq t$ from the production point ($t_0 = 0$).
In Eq.~(\ref{Diracwf})
we discarded the right-handed component $\psi_R$
of the Dirac spinor since it is of order $m/2\kappa$
as compared to the left-handed component,
and we also neglect the positive-helicity
component $x^\prime\phi_{+}$ since it is
of order $\frac{cB}{\kappa}$ relative to $x\phi_{-}$.

We now consider
the possible relevance of the effect of the extra terms
due to the magnetic field on the resonant
flavor conversion.  At each point along the neutrino path, $H_B$
can be diagonalized by the unitary transformation
\begin{equation}\label{Umatter}
U_m (r) =
\left(\begin{array}{cc}
\cos\theta_m& \sin\theta_m\\
-\sin\theta_m &\cos\theta_m
\end{array}\right)\,,
\end{equation}
with the  mixing angle in matter $\theta_m$
determined by
\begin{equation}\label{eq.r43}
\sin2\theta_m =\frac{\Delta_0 \sin 2\theta}{\sqrt{\left(\Delta_0\cos 2\theta  -
v\right )^2
+ \Delta_0^2\sin^2 2\theta}}\,,
\end{equation}
where
\begin {equation}
v(r) = b_e(r) - c_e(r)\vec B\cdot\hat\kappa\,,
\end {equation}
and $\Delta_0 \equiv (m_2^2 - m_1^2)/2\kappa$ .
The denominator in Eq.~(\ref{eq.r43}) corresponds to the difference
between the instantaneous energy eigenvalues of the
neutrino modes $\omega_2 - \omega_1$.
For antineutrinos the $b_e$ and $c_e$ terms change sign
and
\begin{equation}\label{eq.r43b}
\sin2 \bar \theta_m =\frac{\Delta_0 \sin 2\theta}{\sqrt{\left(\Delta_0\cos
2\theta  + v\right )^2
+ \Delta_0^2\sin^2 2\theta}}\,.
\end{equation}

According to the above formulas,
the mixing angle in matter is modified by the neutrino
(antineutrino) interaction with the magnetic field.
As  functions of $v$,  $\sin^2 2 \theta_m$ and $\sin^2 2 \bar \theta_m$
exhibit the characteristic form of a Breit-Wigner
resonance.
For  the neutrinos the resonance condition ($\sin2\theta_m = 1$)  is
\begin{equation}\label{rescond}
v(r_{_R}) =\Delta_0\cos 2\theta \,,
\end{equation}
while for the antineutrinos we have
\begin{equation}\label{resantineut}
v(\overline r_{_R}) = - \Delta_0\cos 2\theta\,.
\end{equation}
The  width of the resonance $\sigma $ is given by the length of
the interval of values of $v(r)$, centered around
$v(r_{_R})$,  such that $\sin^22\theta_m\geq {1 \over 2}$, and similarly
for $\sin^22{\bar \theta}_m$. In both cases,
\begin {equation}
\sigma = 2\Delta_0\sin 2\theta \,.
\end {equation}

In principle, Eqs.~(\ref{rescond}) and (\ref{resantineut})
can be verified simultaneously in a medium
with $B \not = 0$.  This is a novel feature
that contrasts  with the situation without the magnetic
field,  where only neutrinos, but not antineutrinos,
can go through a  resonant region.
For negligible values of $\Delta_0\cos 2\theta$
both conditions reduce to
\begin{equation}\label{approxrescond}
v(r_{_R}) \simeq v(\overline r_{_R}) \simeq 0\,,
\end{equation}
and in such case  the resonance for neutrinos and antineutrinos
occur  within regions which are close.
More precisely,
the  above approximations will be good
to the extent  the differences
$|r_{_R} - {\tilde r}|$
and
$|{\overline r}_{_R} - {\tilde r}|$\,,
where
$v({\tilde r}) = 0$,
are small compared with $r_{_R}$ itself.  In turn,
this implies that
\begin{equation} \label{closecond}
\lambda_{_R} \ll  r_{_R} \,,
\end{equation}
where
\begin{equation}\label{scalev}
\lambda_{_R} \equiv \left|{1 \over v}  \ {dv \over dr} \right|^{-1}_{r =
r_{_R}}
\end{equation}
is the characteristic scale length of $v$ at the neutrino resonance,
and we have taken into account the fact that
$\lambda_{_R} \simeq {1 \over 2} |r_{_R} - {\overline r}_{_R}|$.
It is pertinent  to remark  that  the  condition given by
Eq.~(\ref {closecond}) does not necessarily imply that
the resonance regions of $\nu$ and $\overline \nu$ overlap.
This will happen whenever the separation  
between
the two resonance points is smaller than the average
spatial extension of the  regions.  For the neutrinos, the
actual spatial extension of the resonance region is given by
$\delta_{\nu} = \sigma/|v^\prime(r_R)|\simeq 2\lambda_{_R}\tan 2 \theta$,
where $v^\prime = dv/dr$,
with an analogous
expression for $\delta_{\overline \nu}$.  The condition that
both regions overlap 
is $|r_{_R} - {\overline r}_{_R}| < \frac{1}{2}(\delta_\nu 
+ \delta_{\overline\nu})$, and assuming
$v'(r_{_R}) \simeq v'(\overline r_{_R})$ it
simplifies to  $\tan 2\theta {\
\lower-1.2pt\vbox{\hbox{\rlap{$>$}\lower5pt\vbox{\hbox{$\sim$}}}}\ } 1$\,,
which requires large values of the mixing angle.

If the magnetic fields are
sufficiently strong that Eq.~(\ref{approxrescond}) can be satisfied,
then the phenomenon of resonant oscillations can take
place, without a severe requirement on
the masses and mixing angles of the neutrinos.
Thus, even if the values of these parameters
are constrained by the condition for
resonant oscillations in the Sun and/or
other physical phenomena,
the supernova may simultaneously support resonant
oscillations under the conditions just stated.
In order to estimate the order of magnitude of the
magnetic field needed  to satisfy
Eq.~(\ref{approxrescond}), in what follows we evaluate
$c_e$ in various limiting cases,
which are easily obtained from Eq.~(\ref{bece}).
\begin{description}
\item[Degenerate gas.] In the limit $T\rightarrow 0$, we have
$n_{{\overline e}} = 0$ and
\begin{equation}
\label{degenerate}
c_e = - \sqrt{2}G_Fm_e\mu_B\left(\frac{3n_e}{\pi^4}\right)^{1/3}\,.
\end{equation}
where $\mu_B = e/2m_e$ is the Bohr magneton. Eq.~(\ref{approxrescond})
then requires
\begin{equation}\label{degrescond}
-\mu_B \vec B\cdot\hat\kappa =
\frac{n_e^{2/3}}{m_e}\left(\frac{\pi^4}{3}\right)^{1/3} \,,
\end{equation}
which can be written in the form
\begin{equation}\label{degrescond2}
\left(\frac{\vec B\cdot\hat\kappa}{10^{14} gauss}\right) =
67\left(\frac{Y_e}{0.3}\right)^{2/3}
\left(\frac{\rho}{\rho_0}\right)^{2/3}\,,
\end{equation}
where $\rho$ is the mass density, $Y_e$ is the fractional
number density of electrons and $\rho\equiv 10^{10} g/cm^3$.

\item[Ultrarrelativistic non-degenerate gas.]
In the ultrarelativistic limit, we obtain
\begin{equation}\label{classialur1}
c_e = -\frac{G_Fe}{\sqrt{2}\pi^2}\mu\,.
\end{equation}
Approximating
the Fermi-Dirac distribution by its classical limit
we obtain
\begin{equation}\label{classialur2}
\frac{\pi^2\beta^3}{4}(n_e - n_{{\overline e}})
= \sinh \beta\mu \,.
\end{equation}
Then, for small values of $n_e - n_{{\overline e}}$,
Eq.~(\ref{classialur1}) reduces to
\begin{equation}\label{classialur3}
c_e \simeq -\frac{G_F}{2\sqrt{2}}m_e\mu_B\beta^2(n_e - n_{{\overline e}})\,.
\end{equation}
and the condition in Eq.~(\ref{approxrescond}) translates to
\begin{equation}\label{rescondclassurnum}
\left(\frac{\vec B\cdot\hat\kappa}{10^{14} gauss}\right) =
(1.4\times 10^{3})\left(\frac{T}{10 MeV}\right)^2\,.
\end{equation}

\item[Non-relativistic Boltzmann gas.] For this case we
can borrow the result given in Ref.~\cite{DNP1},
\begin{equation}\label{callsicalnr}
c_e = -\sqrt{2}G_F\mu_B\beta(n_e - n_{{\overline e}})\,,
\end{equation}
and Eq.~(\ref{approxrescond}) then becomes
\begin{equation}\label{rescondclassnrnum}
\left(\frac{\vec B\cdot\hat\kappa}{10^{14} gauss}\right) =
17\left(\frac{T}{10 MeV}\right)\,.
\end{equation}
\end{description}
It is noteworthy that, in the classical limit, the resonance
condition becomes approximately independent of the
number densities.

The estimates given in Eqs.~(\ref{degrescond2}) and (\ref{rescondclassnrnum})
indicate that the phenomenon we have considered may be relevant
in the study of resonant oscillations in the supernova,
where the densities are typically of order $10^{10} g/cm^{3}$,
and magnetic fields of order $10^{14} gauss$ have been
considered.
Our results suggest that the effect
is worthy of further attention and
detailed numerical
studies in order
to asses the range of implications
in a concrete way.

The possibility that the induced neutrino electromagnetic
vertex may contribute to the index of refraction
in the presence
of an external magnetic field
has also been recently pointed
out\cite{kikuchi}.
However, that work did not include the
presence of the matter term ($b_e$)
in the evolution equation, and therefore
did not consider the implications of the
combined effect of that term plus
the magnetic field term ($c_e$)
on the resonant oscillations.

After this work was completed and while this manuscript was being
prepared, we received a preprint\cite{esposito} discussing the same effect
that we consider here.  However, the physical outlook
and the conclusions reached
there regarding the new resonant
condition and its implications are different from ours.

\paragraph*{Acknowledgements~:~}
We wish to acknowledge useful correspondence
with Per Elmfors regarding the results
of Ref.\ \cite{elmfors} and the
contribution of the terms of higher orders in $B$.
The work of JCD was partially supported by Grant No. DGAPA-IN100694 and
that of JFN by the US National
Science Foundation Grant PHY-9320692.

\end{document}